\documentclass[preprint,showpacs,preprintnumbers,amsmath,amssymb]{revtex4}

\usepackage{graphicx}% Include figure files
\usepackage{dcolumn}% Align table columns on decimal point
\usepackage{bm}% bold math

\begin{document}

\preprint{DRAFT 09-10-05}

\title{{\bf Reduced Deadtime and Higher Rate Photon-Counting Detection using a Multiplexed Detector
Array }}

\author{S. Castelletto}
% \email{castelle@ien.it}

\author{I. P. Degiovanni}%
% \email{degio@ien.it}
\author{V. Schettini}

\affiliation{%
Istituto Elettrotecnico Nazionale G. Ferraris \\
Strada delle Cacce 91-10135 Torino (Italy)
}%

\author{A. Migdall}
\affiliation{%
Optical Technology Division\\
National Institute of Standards and Technology, Gaithersburg,
Maryland 20899-8441 }

\date{\today}

% It is always \today, today,
             %  but any date may be explicitly specified

\begin{abstract}
We present a scheme for a photon-counting detection system that can
be operated at incident photon rates higher than otherwise possible
by suppressing the effects of detector deadtime. The method uses an
array of \emph{N} detectors and a 1-by-\emph{N} optical switch with
a control circuit to direct input light to live detectors. Our
calculations and models highlight the advantages of the technique.
In particular, using this scheme, a group of \emph{N} detectors
provides an improvement in operation rate that can exceed the
improvement that would be obtained by a single detector with
deadtime reduced by 1/\emph{N}, even if it were feasible to produce
a single detector with such a large improvement in deadtime. We
model the system for CW and pulsed light sources, both of which are
important for quantum metrology and quantum key distribution
applications.
\end{abstract}

\pacs{Valid PACS appear here}% PACS, the Physics and Astronomy
                             % Classification Scheme.
%\keywords{Suggested keywords}%Use showkeys class option if keyword
                              %display desired
\maketitle

\section{\label{sec:level1}Introduction}
There is a long history of low light level measurement applications,
such as astronomy and particle physics, with demanding detector
requirements. While these applications have provided a steady
motivation for detector improvement, the growing interest and
advancing efforts in quantum information have brought into sharper
focus the need for better photon-counting detectors \cite{MID04}.

Quantum communication and quantum computation applications place
difficult design requirements on the manipulation and processing of
single photons \cite{KKM04}. Quantum cryptography \cite{BEB84} would
particularly benefit from improved detectors, as that application in
the form of Quantum Key Distribution (QKD), is currently
significantly constrained by detector characteristics such as
detection efficiency, dark count rate, timing jitter, and deadtime
\cite{gisinrevmod}. Because of demands for higher-rate secret key
production, the quantum information community is presently engaged
in a number of efforts aimed at improving QKD, including optimizing
the quantum channels for minimum loss \cite{THB02, CFA04}, improving
detector efficiency \cite{KKM04, LZC96, RLM05}, reducing detector
timing jitter \cite{CGL04}, reducing detector deadtime \cite{RBP02},
and single-mode single-photon source engineering \cite{SKK00, BUW01,
USB04, LVS04, RFD05, FAW05, FMW05}. Moreover, with the exponential
growth in multimode parametric downconversion (PDC) photon pair
production rates now in the range of 2x10$^6$ s$^{-1}$  \cite{KAB04}
and the more recent development of $\chi^{(3)}$ single-mode
fiber-based sources with pair rates up to 10$^7$s$^{-1}$
\cite{FAW05,FMW05}, the need is clear for better photon-counting
detection by all means possible, including improved deadtime.

In the area of metrology, high-speed detection capability could
allow the calibration of a very bright single-photon source, that in
the long term could be a viatic for a radiometric standard yielding
a ``quantum candela"\cite{SPW05NPL}. Additional motivation for this
proposal is the improvement of traditional detection applications
such as medical diagnosis, bioluminescence, chemical analysis, and
material analysis \cite{
FinocchiaroSPW2005,CovaSPW2005,ALY02,tudisco}.

This idea is an extension of the already established advantage of
multiplexing many individual, but imperfect, components into a
system that operates with characteristics much closer to the ideal.
An example in the field of single-photon technology is the
multiplexed single-photon PDC source \cite{MBC02}. In electronics, a
more ubiquitous example of this principle would be a computer memory
chip or a disk drive where system control bypasses dead or defective
subunits. This proposal is also becoming more feasible given the
current trends toward integrated microchip arrays of optical sources
and detectors, some of which are now becoming more readily available
\cite{FinocchiaroSPW2005, CovaSPW2005, ALY02}.

 We present a scheme that can
achieve higher detection rates than is otherwise possible by
reducing the effect of detector deadtime. Deadtime is the time
needed after a photon-counting detector fires for the detector to
recover so that it is ready to register a new photon. This recovery
time may be due to the detector, the processing electronics, or some
combination of the two. Photomultipliers are a case where the
detector deadtime contribution can be quite short, and the
subsequent electronics often ultimately set the overall detection
deadtime. In avalanche photodiodes (APDs) though, it is more
difficult to neatly separate deadtime due to ``detection" and
deadtime due to ``electronics." In an APD, the current avalanche
must be physically quenched before the detector is ready for another
photon, resulting in a minimum deadtime of typically a few tens of
ns. In addition, APDs suffer from afterpulsing that requires an
additional wait time before reactivating the detector to avoid a
secondary output pulse caused by the previous photon event. In
typical APD devices, the resulting deadtimes range from $\approx$50
ns for actively quenched APDs, to $\approx$10 $\mu$s for passively
quenched APDs, although even actively quenched APDs sometimes employ
$\mu$s deadtimes to avoid excessive afterpulsing rates. (PMTs also
can suffer from afterpulsing, but modern PMTs typically exhibit
afterpulsing at much lower rates than APDs \cite{RCAPMT}.)

In practice, detectors are usually operated at detection rates much
lower than the inverse of the deadtime to avoid high deadtime
fractions and the associated large deadtime corrections. The
deadtime fraction (DTF) may be defined as the ratio of missed- to
incident-events. Alternately, in the case of a Poissonian CW source,
it may be defined as the fraction of the time the detector spends in
its recovery state (where it is effectively blind to incoming
photons) relative to the total elapsed time. A DTF of 10 \% is often
a reasonable limit for detector operation. The result is that while
many applications would benefit from tens of MHz to GHz detection
rates, the reality is that detectors are in practice limited  to
$\sim$ 1 MHz rates at best. Clearly a way to increase detection
rates is needed.

Our scheme to improve detection rates takes a pool of
photon-counting detectors and operates them as a unit, or a
``detection resource," in a way that allows overall detection at
higher rates than would be possible if the detectors were operated
individually, while maintaining comparable DTFs. We model and
numerically analyze the scheme for typical detector deadtimes, and
show the superiority of the scheme over hypothetical single
detectors with much improved deadtimes both for CW- and
pulsed-sources. We also note that our scheme bears some resemblance
to schemes using beamsplitter trees and detector arrays \cite{KOB01,
RHH03, PTK96}. We compare the proposed scheme DTFs to those tree
schemes, as well as to the performance of a single detector with
much reduced deadtime.

\section{Scheme}

The detection scheme relies on the rather obvious fact that, while a
detector has a significant deadtime when it does fire, it has no
deadtime when it does not fire. The scheme consists of a
1-by-\emph{N} optical switch that takes a single input stream of
photons and distributes them to members of an array of \emph{N}
detectors. A switch control circuit monitors which detectors have
fired recently and are thus dead, and then routes subsequent
incoming pulses to a detector that is ready. This system allows a
system of \emph{N} detectors to be operated at a significantly
higher detection rate than \emph{N} times the detection rate of an
individual detector, while maintaining the same DTF.

To understand the process, consider a fixed input photon pulse rate,
some pulses of which may contain a photon and some may not. (For
example, this is usually the situation in a quantum cryptography
application.) At the start of operation, all detectors are live and
ready to detect a photon. The switch is set to direct the first
incoming pulse to the first detector of the array. Control
electronics monitor the output of that detector to determine if it
fires. If the detector does fire, the control switches the next
pulse to the next detector. If the detector does not fire, then the
switch state remains unchanged. The process repeats with the input
always directed to the first available live detector. At high count
rates many of the detectors may fire in a short period of time and
subsequently be in their dead state, but as long at the first
detector recovers to its live state before the last detector
triggers, the system will still be live and ready to register an
incoming photon. The system will only be dead when all detectors
have fired within one deadtime of each other. The system operation
could be sequential with each detector firing in order as just
described, or it could be set up to direct the input to any live
detector. That would allow for optimum use of an array of detectors
where each detector may have a different deadtime.

\begin{figure}[tbp]
%[htbp]
\par
\begin{center}
\includegraphics[angle=0, width=12 cm]{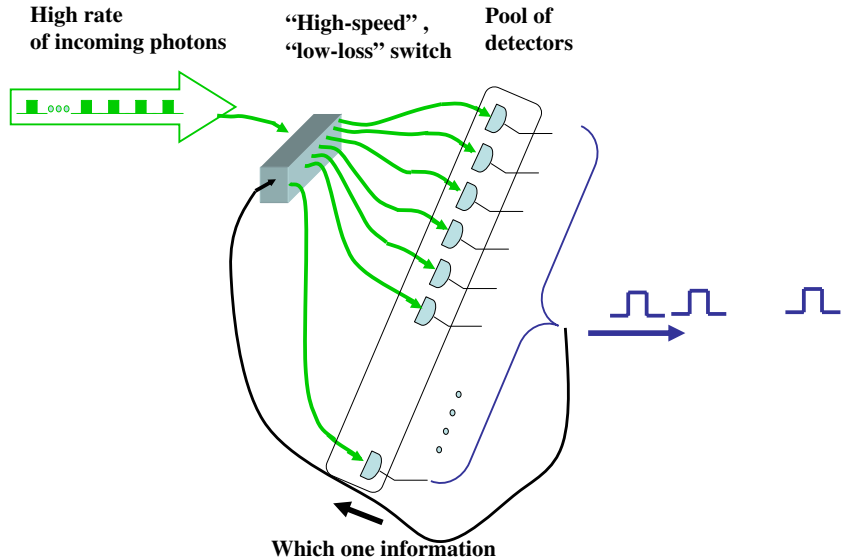}
\end{center}
\caption{ A pool of detectors and a fast switch are used to register
a high rate of incoming photons. Incoming photons are switched to a
ready detector. If it fires, the detector is switched out of the
ready pool until recovery. If it does not fire, that detector
remains ready. } \label{Figure 2}
\end{figure}

\section{\label{sec:level1}Analysis and models}

To understand the system operation and quantify its advantages
relative to non-multiplexed systems, we use approximate analytical
and numerical Monte Carlo models, both for the cases of a CW Poisson
distributed- and a pulsed-source.

\subsection{\label{sec:level1}CW (Poisson) source: analytical modeling}

Our analytical calculation estimates the DTF from the mean total
count rate to the overall detector pool and effective deadtimes for
each detector (which depend on their position in the switching
system). We consider a Poissonian source and a pool of detectors
with the identical detection efficiencies $\eta$ and identical
non-extending deadtimes $T_\text{d}$ \cite{metrologia}. Zero switch
transition time is assumed. (We refer to a Poissonian source as CW
because, while the photons arrive at discrete times, they have equal
probability to arrive at any time.) Furthermore, the optical switch
is programmed to send photons to the detectors in sequence. The
switch sends photons always to detector 1 ($\text{D}_1$) when it is
live. If it is dead, it sends the photons to $\text{D}_2$, and so
on.

The probability that $n$ photons from a Poissonian source with mean
photon rate $\lambda$ are registered by a single live detector with
efficiency $\eta$ in a time interval $T$ is $P(n)= (\eta \lambda
T)^{n} e^{-\eta \lambda T} / n!$. Thus, the mean number of counts
registered is $\eta \lambda T$. From here on for simplicity we
assume $\eta=1$. (We can do this without loss of generality, as
$\eta$ and $\lambda$ always appear together and can thus be traded
off against each other with affecting the ultimate results.) In the
presence of deadtime $T_\text{d}$ and for measurement time $T \gg
T_\text{d}$, the mean number of counts registered reduces to
\begin{equation} \label{eq1}
M=  \lambda T -M \lambda T_\text{d},
\end{equation}
where $M\lambda T_\text{d}$ accounts for the mean number of photons
lost to deadtime. Rearranging, we have
\begin{equation} \label{eq2}
M = \frac{ \lambda T}{1+ \lambda T_\text{d}}.
\end{equation}
 The DTF, defined
as the ratio of the lost counts over the total counts in the
absence of dead time, for this simple case is
\begin{equation} \label{DTF}
\mathrm{DTF} = \frac{ \lambda T- M }{\lambda T} = 1- \frac{ 1}{1+
\lambda T_\text{d}}.
\end{equation}

Now consider an array of detectors connected to an optical switch
(Fig. 1) with switching time negligible with respect to
$T_\text{d}$. Eq. (\ref{eq2}) holds for $\text{D}_1$, so the mean
number of counts detected by $\text{D}_1$ is $M_{1} = \frac{ \lambda
T}{1+ \lambda T_\text{d}}$, while $\text{D}_1$ is dead during a time
interval $T_{2}= M_{1} T_\text{d}$. Thus the time interval during
which $\text{D}_2$ may count photons is $T_{2}$ \footnote{We
emphasize that despite the fact that the measurement time $T_{2}$ is
composed of separated time intervals, while $T_{1}$ is a single time
interval, the source is Poissonian and this allows us to consider
$T_{2}$ as a continuous time interval in the evaluation of this
approximated mean count rate.}. Continuing the analogy with Eq.
(\ref{eq2}) the mean number of counts detected by $\text{D}_2$ is
\begin{equation} \label{eq2bi}
M_{2}  = \frac{ \lambda T_{2}}{1+ \lambda
\mathcal{T}_{\text{d}(2)}},
\end{equation}
where here $ \mathcal{T}_{\text{d}(2)}$ is the effective deadtime
associated with $\text{D}_2$. It is necessary to introduce an
effective deadtime because the measurement time $T_{2}$ is not
reduced by the full deadtime $T_{\text{d}}$. Only part of the
deadtime of $\text{D}_2$ will occur while $\text{D}_1$ is dead,
effectively reducing $\mathcal{T}_{\text{d}(2)}$. We postpone the
evaluation of effective deadtimes.

In analogy with the arguments leading to Eq. (\ref{eq2}),
$\text{D}_3$ is live during the time interval $T_{3}=M_{2}
\mathcal{T}_{\text{d}(2)}$, corresponding to the time interval when
both $\text{D}_1$ and $\text{D}_2$ are dead, and the mean number of
counts registered by $\text{D}_3$ is
\begin{equation} \label{eq3}
M_{3} = \frac{ \lambda T_{3}}{1+ \lambda \mathcal{T}_{\text{d}(3)}},
\end{equation}
where $ \mathcal{T}_{\text{d}(3)}$ is the effective deadtime
associated with $\text{D}_3$. Likewise for detector $\text{D}_i$,
the measurement time is $T_{i}=M_{i-1} \mathcal{T}_{\text{d}(i-1)}$
and the mean count rate is $M_{i}  = \frac{ \lambda T_{i}}{1+
\lambda \mathcal{T}_{\text{d}(i)}}$.

The mean number of counts registered by the multiplexed detector
system with N-detectors is $M_{\text{tot}}=M_{1}+ M_{2} +
...+M_{N}$, and the overall system DTF $=\frac{\lambda T -
M_{\text{tot}}}{\lambda T}$ is
\begin{eqnarray} \label{eq3}
\mathrm{DTF}&=&1- \frac{ 1}{1+ \lambda T_\text{d}}\left(
\begin{array} {c} 1 + \frac{ \lambda T_\text{d}}{1+\lambda
\mathcal{T}_{\text{d}(2)}}+ \frac{ \lambda^{2} T_\text{d}
\mathcal{T}_{\text{d}(2)} }{(1+ \lambda
\mathcal{T}_{\text{d}(2)})(1+ \lambda \mathcal{T}_{\text{d}(3)})} +
...\\  + \frac{ \lambda^{N-1} T_\text{d}
\mathcal{T}_{\text{d}(2)}...\mathcal{T}_{\text{d}(N-1)} }{(1+
\lambda \mathcal{T}_{\text{d}(2)})(1+ \lambda
\mathcal{T}_{\text{d}(3)})...(1+ \lambda
\mathcal{T}_{\text{d}(N)})}\end{array} \right).
\end{eqnarray}

 For comparison, as we will see in our subsequent analysis and modeling, the DTF obtainable by simply reducing
$T_\text{d}$ of a single detector by a factor of $1/N$ is
DTF=$1-\frac{1}{1+ \lambda T_\text{d}/N}$. We note also that this
result is the same as would be obtained by an array of \emph{N}
detectors with deadtime $T_\text{d}$ and passive switching such as
may be implemented with a tree arrangement of beam splitters. This
result follows from the fact that, in such a tree, the incident rate
at each detector is $\lambda/N$.

We analyse the effective deadtime of $\text{D}_2$ for two cases
using Fig. 2. The top timeline indicates the arrival times of
photons. The 2nd and 3rd timelines indicate when $\text{D}_1$ and
$\text{D}_2$ register counts and when they are dead (shaded
regions). $\Delta =t_{3}-t_{1}$ is the time interval between the
first photon counted by $\text{D}_1$ and the first one after its
deadtime $T_\text{d}$ has expired. $\delta =t_{2}- t_{1}$ is the
time interval between the first two photon arrivals at times $t_{1}$
and $t_{2}$, the first detected by $\text{D}_1$ and the second by
$\text{D}_2$. Fig. 2 shows two possible situations. In case (a), the
time interval between two sequential counts of $\text{D}_1$ ($t_1$,
$t_3$) is larger than the time interval between the count at $t_1$
and the count at $t_2$ plus $T_{\text{d}}$ the deadtime of
$\text{D}_1$, namely $\Delta
> \delta  + T_{\text{d}}$. In this case the effective deadtime of the detector
combination of $\text{D}_1$ and $\text{D}_2$,
$\mathcal{T}_{\text{d}(2)}$, is $T_\text{d}-\delta $. In case (b)
the time interval $\Delta $ is shorter than $\delta + T_\text{d}$,
thus two terms contribute to the effective deadtime,
$\mathcal{T}_{1}=T_\text{d}-\delta $ and $\mathcal{T}_{2}=
T_\text{d} + \delta
 - \Delta $.

\begin{figure}[tbp]
%[htbp]
\par
\begin{center}
\includegraphics[angle=0, width=12 cm]{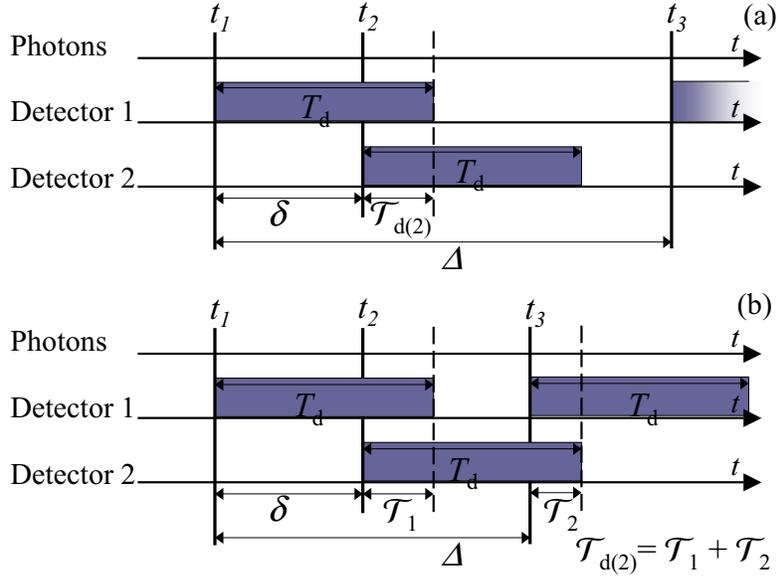}
\end{center}
\caption{ The effective deadtime of $\text{D}_1$. Case (a) the time
interval between two subsequent counts of $\text{D}_1$, $\Delta >
\delta+ T_\text{d} $,with $\delta$ the time interval between the
count of $\text{D}_1$ and the subsequent photon counted by
$\text{D}_2$ during the deadtime of $\text{D}_1$, and $T_\text{d}$
the single detector deadtime. Case (b) $\Delta < \delta+ T_\text{d}
$. In this second case two terms, $\mathcal{T}_{1}$ and
$\mathcal{T}_{2}$ contribute to the final deadtime. Dark shaded
regions represent the deadtime of the individual detectors}
\label{Figure 2}
\end{figure}

As we assumed that the arrival of photons at the array of
detectors is Poissonian, the random variables $\Delta $ and
$\delta $ are statistically independent. The probability density
function of the random variable $\Delta$ is $f_{\Delta}(\Delta )=
\lambda e^{- \lambda (\Delta - T_\text{d})} \Theta(\Delta -
T_\text{d}) $, where $\Theta(x)=1$ for $x>0$, and 0 otherwise. The
probability density function of the random variable $\delta$ is
$f_{\delta}(\delta
 )=  \lambda e^{-  \lambda \delta }/(1-e^{-  \lambda T_\text{d}})
\Theta(T_\text{d}-\delta ) $.

The probability that situation (a) occurs is
\begin{equation} \label{eqpa}
p_{\text{a}}=\int_{\Delta  >\delta   + T_\text{d}} f_{\Delta}(\Delta
) f_{\delta}(\delta  ) \mathrm{d} \Delta   \mathrm{d} \delta  ,
\end{equation}
while the probability that situation (b) occurs is
\begin{equation} \label{eqpb}
p_{\text{b}}=\int_{\Delta  <\delta   + T_\text{d}} f_{\Delta}(\Delta
) f_{\delta}(\delta  ) \mathrm{d} \Delta   \mathrm{d} \delta  .
\end{equation}

In case (a), the mean value of the ``effective" deadtime is
$\mathcal{T}_{\text{a}} = T_\text{d} - \mathrm{E}_{\text{a}}(\delta
)$ with
\begin{equation} \label{eqda}
\mathrm{E}_{\text{a}}(\delta  )=\frac{\int_{\Delta  >\delta  +
T_\text{d}} \delta
  f_{\Delta}(\Delta  ) f_{\delta}(\delta  ) \mathrm{d} \Delta
\mathrm{d} \delta  }{\int_{\Delta >\delta   + T_\text{d}}
f_{\Delta}(\Delta
 ) f_{\delta}(\delta  ) \mathrm{d} \Delta   \mathrm{d} \delta  }.
\end{equation}

In case (b), the mean value of the ``effective" deadtime is
$\mathcal{T}_{\text{b}} = 2 T_\text{d} -
\mathrm{E}_{\text{b}}(\Delta  )$ with
\begin{equation} \label{eqdb}
\mathrm{E}_{\text{b}}(\Delta  )=\frac{\int_{\Delta  <\delta   +
T_\text{d}} \Delta
  f_{\Delta}(\Delta  ) f_{\delta}(\delta  ) \mathrm{d} \Delta
\mathrm{d} \delta  }{\int_{\Delta  <\delta   + T_\text{d}}
f_{\Delta}(\Delta ) f_{\delta}(\delta  ) \mathrm{d} \Delta
\mathrm{d} \delta  }.
\end{equation}

The mean effective deadtime $\mathcal{T}_{\text{d}(2)}= p_{\text{a}}
\mathcal{T}_{\text{a}} + p_{\text{b}} \mathcal{T}_{\text{b}}$ can be
calculated as
\begin{equation} \label{fd}
\mathcal{T}_{\text{d}(2)}= T_\text{d} -\frac{1-e^{- \lambda
T_\text{d}}}{2 \lambda}.
\end{equation}

 We iterate the formula for the subsequent
detectors obtaining a recursive expression for the effective final
deadtime, $\mathcal{T}_{\text{d}(i)}= \mathcal{T}_{\text{d}(i-1)}
-\frac{1-e^{- \lambda \mathcal{T}_{\text{d}(i-1)}}}{2 \lambda}$. The
calculated results for DTF versus incident photon rate for pools of
up to 12 detectors are nearly identical to the Monte Carlo results
shown in Fig. 3 and described in the next section.

\subsection{CW Source: Monte Carlo Modeling}

The Monte Carlo model assumes a CW source with Poisson distributed
incident photons at a range of rates meant to describe the use of
the system in conjunction with a laser source. As mentioned before,
we can assume 100~\% efficient collection and detection without loss
of generality. The individual detector deadtimes were set to 50 ns.
The modeling procedure consisted of first using a random number
generator to simulate an input stream of a large number of photons
with Poisson distributed arrival times. The resulting photon list
was then apportioned to the first detector by going through each
photon time on the list in sequence to see if it could have been
detected by a single detector $\text{D}_1$. That is, once a photon
is detected, any photons within one deadtime after that detected
photon are skipped. A new  list consisting of the ``skipped" photons
was then apportioned to the second detector $\text{D}_2$ using the
same procedure as for the first detector. This process was repeated
for all \emph{N} detectors. Those photons left after detector
$\text{D}_N$ are those that would be missed by the system and the
ratio to the total number in the original photon list is the
deadtime fraction, as previously defined.

Figure 3a shows the resulting DTF versus mean incident photon rate
for systems of varying numbers of detectors. The Monte Carlo results
(shown) and the analytical calculations (not shown) provide nearly
identical results. From the $R_{\text{DTF}=10\%} $ points (defined
as the incident photon rate at which the DTF=~$10~\%$), we see that
for example, a system of 6 detectors can operate at 32 times the
incident rate of a single detector while maintaining 10~\% deadtime.
This is significantly more than just six times the single detector
count rate, the improvement possible with a passive switch
arrangement,  highlighting the power of the technique. We also see
that the multiplexed detector scheme also has an advantage over
simply reducing the deadtime of an individual detector. Even
reducing the deadtime by a factor of 10 (to 5 ns) does not allow
improvements equal to the system with 6 detectors with
$T_{\text{d}}$ = 50 ns.

Figure 3b shows the dependence of $R_{\text{DTF}=10\%} $  on the
number of detectors in the system, which fits well to a 2nd order
polynomial. The origin of this behavior is due to the correction in
the effective deadtime of each detector of the pool, embedding a
nonlinear dependence on the number of detectors, while the same
behavior is not present for a passive switch (or a tree of
beamsplitters), as shown by its linear $R_{\text{DTF}=10\%}$
dependence on the number of detectors.

For complete modeling of this scheme other switch parameters such as
switch losses, switching transition times, switch latency, maximum
switching rates, and cross-talk should be included. Of these, switch
loss is probably the most problematic, as commercially available ns
switches have losses $\approx$2-3 dB, although there are ongoing
efforts to address this issue. Loss affects the overall detection
efficiency so it should not affect the functional behavior of the
results presented here. Switch transition and latency times should
have effects similar to increasing the deadtime of the individual
detectors as well as reducing overall detection efficiency. However,
with some commercial switch transition times being below 50 ps, that
should not be a severe limit. These parameters will be the subject
of further modeling as they will be what ultimately limits how far
this method can be pushed.
\begin{figure}[tbp]
%[htbp]
\par
\begin{center}
\includegraphics[angle=0,width=15 cm]{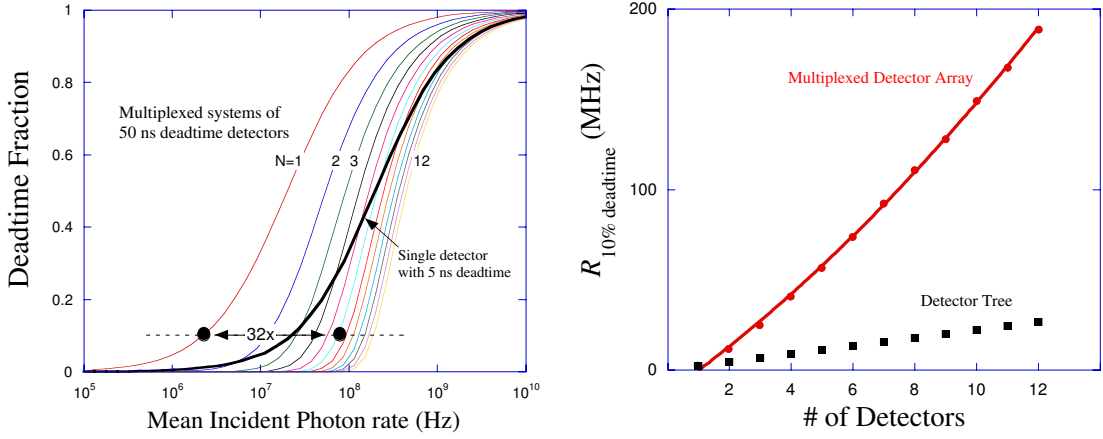}
\end{center}
\caption{a.) DTF versus mean incident photon rate shown for systems
of 1 to 12 detectors, all with 50 ns deadtimes. The dashed
horizontal line shows the 10 \% DTF level that is often the
practical limit for detector operation. The DTF for a single
detector with 5 ns deadtime is shown for comparison (thick line). As
an example of the advantage of the scheme, the two points
illustrating that with 6 detectors the system would be able to
operate at a 32 times larger incident photon rate than a single
detector, while maintaining that 10~\% DTF. b.) $R_{\text{DTF}=10\%}
$ (circles) versus number of detectors in the detector pool, along
with a fit to a 2nd order polynomial (line). For comparison the
dependence of $R_{\text{DTF}=10\%} $ for a detector tree is also
shown (squares). } \label{Figure 3alan}
\end{figure}

%%%%%%%%%%%%%%%%%%%%%%%%%%%%%%%%%%%%%%%%%%%%%%%%%%%%%%%%%%%%%%%%%%%%%%%%%%%%%%%%%%%%%%%%%%%%%%%%%%%%%%%%%
%%%%%%%%%%%%%%%%%%%%%%%%%%%%%%%%%%%%%%%%%%%%%%%%%%%%%%%%%%%%%%%%%%%%%%%%%%%%%%%%%%%%%%%%%%%%%%%%%%%%%%%%%%%

\subsection{\label{sec:level1}Pulsed source: analytical modeling}

The detection scheme (Fig. 1) modeling with a pulsed source proceeds
in similar fashion to the CW case, and the formulas derived below
are analogous to the CW results. Each pulse of the source may
contain 0, 1 or more photons, but because the detector we are
modeling cannot distinguish between 1 or more photons it has only
two output possibilities: it either fires or it does not. The
probability that a live detector produces a count (``event'') for an
individual pulse is $p$. The probability that $n$ events are counted
by a single live detector in a sequence of $\mathcal{N}=\nu T$
pulses (where $\nu $ is the repetition rate of the pulsed source,
and $T$ the measurement time) is $B(n|\mathcal{N},p)=
\mathcal{N}![n! (\mathcal{N}-n)!]^{-1} p^n (1-p)^{\mathcal{N}-n}$.
From this, it can be shown that the mean number of counts is $p
\mathcal{N}$ and the mean count rate is $p \nu$. In the presence of
deadtime, the detector is dead for a certain number of pulses
$\mathcal{N}_\text{d}=\mathrm{Int}(\nu T_{\mathrm{d}})$, where $
T_{\mathrm{d}}$ is the deadtime of the single detector and Int
indicates the integer part. For measurement time such that
$\mathcal{N} \gg \mathcal{N}_\text{d}$, the mean number of counts
reduces to
\begin{equation} \label{eq1pulsed}
M=  p \mathcal{N}  - M ~ p \mathcal{N}_\text{d},
\end{equation}
where $p \mathcal{N}_\text{d} $ is the mean number of events lost
during one deadtime. Thus, the mean number of counted events is
\begin{equation} \label{eq2pulsed}
M = \frac{ p \mathcal{N}}{1+ p \mathcal{N}_\text{d}}.
\end{equation}

Eq. (\ref{eq2pulsed}) holds for the first detector $\text{D}_1$, so
the mean number of counts detected by $\text{D}_1$ is $M_{1} =
\frac{ p \mathcal{N}}{1+ p \mathcal{N}_\text{d}}$, while
$\text{D}_1$ is dead for the average number of pulses in the
measurement time $\mathcal{N}_{2}= M_{1} \mathcal{N}_\text{d}$.
Thus, the time interval during which $\text{D}_2$ may count photons
is $\mathcal{N}_{2}$. Continuing the analogy with the CW case, the
mean number of counts detected by $\text{D}_2$ is
\begin{equation} \label{eq2bis}
M_{2}  = \frac{ p \mathcal{N}_{2}} {1+ p
\mathcal{N}_{\text{d}{(2)}}},
\end{equation}
where here $ \mathcal{N}_{\text{d}(2)}$ is the mean number of pulses
constituting this ``effective deadtime'' associated to $\text{D}_2$.
The evaluation of the ``effective deadtimes'' is presented below.
Thus, for $\text{D}_i$, the average number of pulses in the
measurement time is $\mathcal{N}_{(i)}=M_{i-1}
\mathcal{N}_{\text{d}(i-1)}$, and the mean count rate is $M_{i}  =
\frac{ p \mathcal{N}_{i}}{1+ p \mathcal{N}_{\text{d}(i)}}$.

The mean number of counts registered by the multiplexed detector
system is $M_{\text{tot}}=M_{1}+M_{2}+...+ M_{N}$, and the
DTF$=\frac{p \mathcal{N}-M_{\text{tot}}}{p \mathcal{N}}$ is
\begin{eqnarray} \label{eq15}
\mathrm{DTF}= 1- \frac{1}{1+ p \mathcal{N}_\text{d}}\left(
\begin{array} {c} 1 + \frac{ p \mathcal{N}_{\text{d}}}{1+p
\mathcal{N}_{\text{d}{(2)}}}+ \frac{ p^{2} \mathcal{N}_{\text{d}}
\mathcal{N}_{\text{d}(2)} }{(1+ p \mathcal{N}_{\text{d}(2)})(1+ p
\mathcal{N}_{\text{d}(3)})} + ...\\ \nonumber + \frac{ p^{N-1}
\mathcal{N}_{\text{d}} \mathcal{N}_{\text{d}(2)}
...\mathcal{N}_{\text{d}(N-1)}}{(1+ p
\mathcal{N}_{\text{d}(2)})(1+ p \mathcal{N}_{\text{d}(3)})...(1+ p
\mathcal{N}_{\text{d}(N)})}\end{array}\right).
\end{eqnarray}
 We note that if the number of the detectors in the multiplexed
array is more than $ \mathcal{N}_\text{d}+1 $, all the events will
be detected and DTF will be always zero.

In the pulsed case, the advantage obtainable with a single
detector with deadtime reduced of a factor 1/$N$ is given by
DTF$=1-\frac{1}{1+ p ~ \mathrm{Int}(\nu T_\text{d}/N)}$, while for
the detector tree configuration the deadtime fraction is
DTF$=1-\frac{1}{1+ p
 \mathcal{N}_{\mathrm{d}} /N}$.
Note that for the case of a single detector with reduced deadtime,
when $T_{\text{d}}/N < {1}/{\nu}$ then the DTF=0, while this is
never the case for the detector tree configuration.

%%%%%%%%%%%%%%%%%%%%%%%%%%%%%%%%%%%%%%%%%%%%%%%%%%%%%%%%%%%%%%%%%%%%%%%%%

We analyze the ``effective" deadtime of $\text{D}_2$ for a pulsed
source using Fig. 4 where the dashed vertical lines represent
empty pulses and the continuous vertical lines represent detection
events. As with Fig. 2, the $\text{D}_1$ and $\text{D}_2$
timelines indicate when $\text{D}_1$ and $\text{D}_2$ register a
count and when they are dead (dark shaded regions). $n_{\Delta}$
is the number of pulses between the first photon counted by
$\text{D}_1$ and the subsequent one, after its deadtime
$\mathcal{N}_\text{d}$. $n_{\delta}$ is the number of pulses
between the first two events, the first detected by $\text{D}_1$
and the second by $\text{D}_2$. Fig. 5 shows two possible
situations. In case (a), the time interval between two subsequent
counts of $\text{D}_1$ is larger than the time interval between
the first detected by $\text{D}_1$ and the second by $\text{D}_2$
(during the deadtime of $\text{D}_1$) plus the deadtime of the
$\text{D}_2$, namely $n_{\Delta} \geq n_{\delta} +
\mathcal{N}_\text{d}$. In this case, the effective deadtime of the
detector combination of $\text{D}_1$ and $\text{D}_2$,
$\mathcal{N}_{\text{d}(2)}$, is $\mathcal{N}_\text{d}-n_{\delta}
$. In case (b) the time interval $n_{\Delta} $ is shorter then
$n_{\delta} + \mathcal{N}_\text{d}$, thus two terms contribute to
the effective deadtime, the effective deadtime,
$\mathcal{P}_{1}=\mathcal{N}_\text{d}-n_{\delta} $ and
$\mathcal{P}_{2}= \mathcal{N}_\text{d}+ n_{\delta} - n_{\Delta} $.

\begin{figure}[tbp]
%[htbp]
\par
\begin{center}
\includegraphics[angle=0, width=12 cm, height=8 cm]{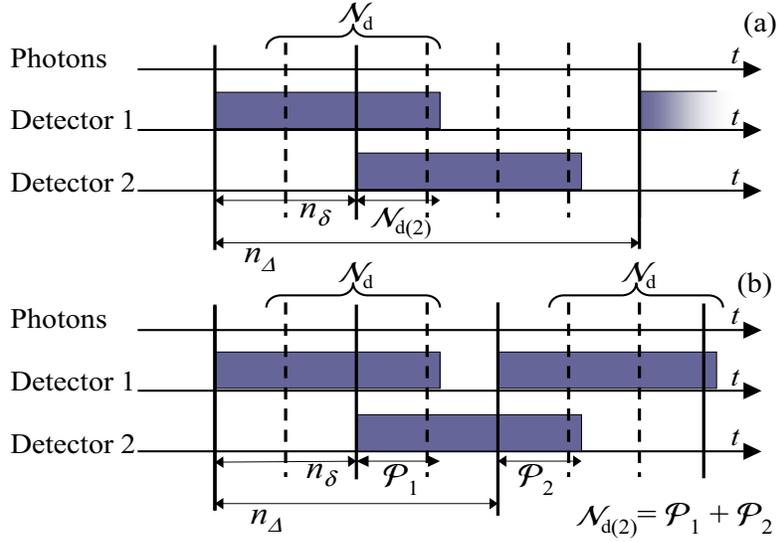}
\end{center}
\caption{ The effective deadtime of $\text{D}_1$. Case (a) the time
interval between two subsequent counts of $\text{D}_1$, $n_{\Delta}
\geq n_{\delta}+ \mathcal{N}_\text{d} $,with $n_{\delta}$ the time
interval between the count of $\text{D}_1$ and the subsequent photon
counted by $\text{D}_2$ during the deadtime of $\text{D}_1$, and
$T_{\text{d}}$ the actual single detector deadtime. Case (b)
$n_{\Delta} < n_{\delta}+ \mathcal{N}_\text{d} $. In this second,
case two terms, $\mathcal{P}_{1}$ and $\mathcal{P}_{2}$ contribute
to the final deadtime. } \label{Figure 2}
\end{figure}

We consider the random variables $n_{\Delta} $ and $n_{\delta} $ to
be statistically independent as there is no correlation between
pulses. The probability distribution of $n_{\Delta}$ is
$P_{\Delta}(n_{\Delta} )= p(1-p)^{n_{\Delta}-\mathcal{N}_\text{d}-1}
$, with  $n_{\Delta}$ integer and  $n_{\Delta}\geq
\mathcal{N}_\text{d}+1 $. The $n_{\delta} $ probability distribution
is $P_{\delta}(n_{\delta}
 )=p(1-p)^{n_{\delta}-1}[1-(1-p)^{\mathcal{N}_\text{d}}]^{-1}  $, with
$n_{\delta}$ integer and $1 \leq n_{\delta}\leq \mathcal{N}_\text{d}
$.

The probability that situation (a) ($n_{\Delta}  \geq n_{\delta} +
\mathcal{N}_\text{d}$) occurs is
\begin{equation} \label{eqpapulsed}
p_{a}=\sum_{n_{\delta}=1  }^{\mathcal{N}_\text{d}}
\sum_{n_{\Delta}=n_{\delta}+\mathcal{N}_\text{d} }^{+\infty}
P_{\Delta}(n_{\Delta} ) P_{\delta}(n_{\delta}  ) ,
\end{equation}
while the probability that situation (b) ($n_{\Delta}  <
n_{\delta} + \mathcal{N}_\text{d}$) occurs is
\begin{equation} \label{eqpbpulsed}
p_{b}=\sum_{n_{\delta}=2  }^{\mathcal{N}_\text{d}}
\sum_{n_{\Delta}=\mathcal{N}_\text{d}+1
}^{n_{\delta}+\mathcal{N}_\text{d}-1} P_{\Delta}(n_{\Delta} )
P_{\delta}(n_{\delta}  ) .
\end{equation}

In case (a) the mean value of the ``effective" the deadtime is
$\mathcal{N}_\text{d,a} = \mathcal{N}_\text{d}  -
E_{\text{a}}(n_{\delta} )$, with
\begin{equation} \label{eqdapulsed}
E_{\text{a}}(n_{\delta}  )=\frac{\sum_{n_{\delta}=1
}^{\mathcal{N}_\text{d}}
\sum_{n_{\Delta}=n_{\delta}+\mathcal{N}_\text{d} }^{+\infty}
n_{\delta} P_{\Delta}(n_{\Delta} ) P_{\delta}(n_{\delta} )
}{\sum_{n_{\delta}=1 }^{\mathcal{N}_\text{d}}
\sum_{n_{\Delta}=n_{\delta}+\mathcal{N}_\text{d} }^{+\infty}
P_{\Delta}(n_{\Delta} ) P_{\delta}(n_{\delta}  ) }.
\end{equation}

While in case (b) the mean value of the ``effective" deadtime is
$\mathcal{N}_\text{d,b} = 2 \mathcal{N}_\text{d} -
E_{\text{b}}(n_{\Delta} )$, with
\begin{equation} \label{eqdbpulsed}
E_{\text{b}}(n_{\Delta}  )=\frac{\sum_{n_{\delta}=2
}^{\mathcal{N}_\text{d}} \sum_{n_{\Delta}=\mathcal{N}_\text{d}
}^{n_{\delta}+\mathcal{N}_\text{d}-1} n_{\Delta}
P_{\Delta}(n_{\Delta} ) P_{\delta}(n_{\delta}  )
}{\sum_{n_{\delta}=2  }^{\mathcal{N}_\text{d}}
\sum_{n_{\Delta}=\mathcal{N}_\text{d}}
^{{n_{\delta}=\mathcal{N}_\text{d} -1}}  {P_{\Delta}(n_{\Delta} )
P_{\delta}(n_{\delta}  )  } }.
\end{equation}

The mean effective deadtime $\mathcal{N}_\text{d(2)}= p_{\text{a}}
\mathcal{N}_\text{d,a} + p_{\text{b}} \mathcal{N}_\text{d,b}$ can be
calculated as
\begin{equation} \label{fkpulsed}
\mathcal{N}_{\text{d}(2)}=
\mathcal{N}_\text{d}-\frac{1-(1-p)^{\mathcal{N}_\text{d}+1}}{(2-p)p}.
\end{equation}

We iterate the formula for the following detectors obtaining a
recursive expression for the effective final deadtime,
$\mathcal{N}_{\text{d}(i)}=
\mathcal{N}_{\text{d}(i-1)}-\frac{1-(1-p)^{\mathcal{N}_{\text{d}(i-1)}+1}}{(2-p)p}$.
The calculated results for DTF versus incident photon rate for pools
of up to 5 detectors are shown in Fig. 5. Monte Carlo results as
described in the next section are also shown.

\begin{figure}[tbp]
%[htbp]
\par
\begin{center}
\includegraphics[angle=0, width=15 cm]{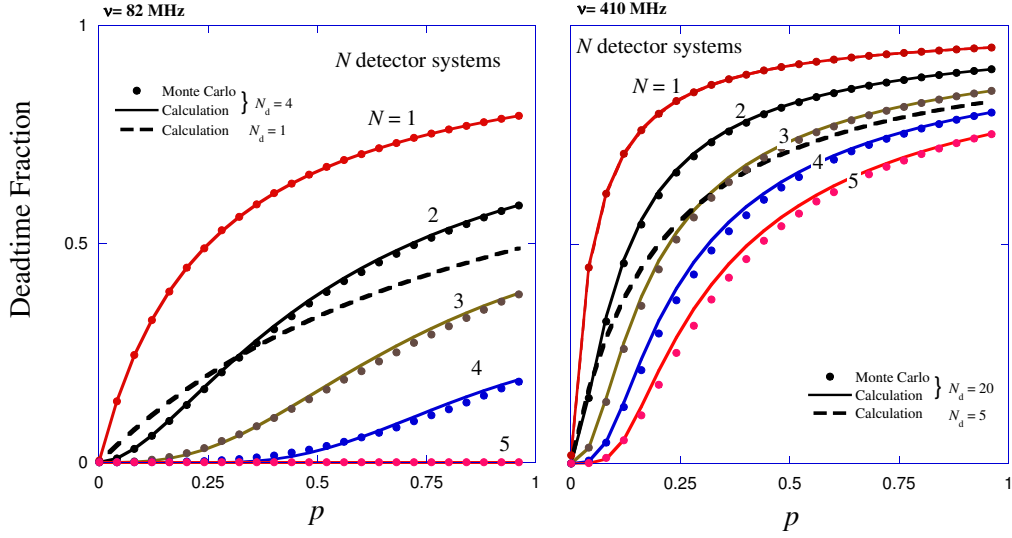}
\end{center}
\caption{Deadtime Fraction versus probability of detection per pulse
for pools of up to 5 detectors, for detectors each with
$T_\text{d}=$ 50 ns, with a pulsed source with repetition rate of
(a) 82 MHz ($\mathcal{N}_\text{d}=4$), and (b) 410 MHz
($\mathcal{N}_\text{d}=20$). Circles represent the Monte Carlo
simulations, while solid lines are the calculated values according
to the here presented theory. For comparison, a single-detector-DTF
(broken line) with a reduced deadtime of 12.5 ns, corresponding to
(a) $\mathcal{N}_\text{d}=1$ and (b) $\mathcal{N}_\text{d}=5$ are
shown.  } \label{Figure 6}
\end{figure}

\subsection{Pulsed Source: Monte Carlo Modeling}

The Monte Carlo model assumes a geometric distribution which samples
pulse arrival times as shown in Appendix A. As mentioned before, the
detectors cannot discriminate between one or more photons in a
single pulse and can fire at most once during a pulse, so the
probability of detecting an event per pulse spans from 0 to 1. As in
the CW case, a random number generator was used to simulate the
source, although this time it is a pulsed source with geometric
distributed events. As before, the resulting event list is
apportioned to each detector in sequence to see if it could have
been detected or if it
is skipped by that detector. %The list of the ``skipped" events is
%then apportioned to the second detector using the same procedure as
%for the first detector, and the process is repeated for all N
%detectors.
Those events left after the $N^{th}$ detector are those that would
be missed by the system and the ratio to the total number in the
original event list is the deadtime fraction.

In Fig. 5 we show the DTF versus the probability of detection of an
event per pulse $p$, for pools of up to 5 detectors each with 50 ns
deadtime, for a pulsed source with repetition rate of (a) 82 MHz
($\mathcal{N}_\text{d}=4$) and of (b) 410 MHz (
$\mathcal{N}_\text{d}=20$). In Fig. 6 (a) we observe that, as
expected, the DTF is 0 in the case of a pool of five or more
detectors.

To highlight the advantage of the multiplexed detector scheme, we
compare the performance of the multiplexed detector system to a
single detector with 4x reduced deadtime of 12.5 ns. Fig. 6, also
shows a comparison of $R_{DTF=10\%}$, for a pulsed source with
repetition rates of 82 MHz and 410 MHz,  for a single detector
with reduced dead time, for the detector tree configuration and
our scheme. Because $R_{DTF=10\%}$ increases quadratically with
the number of detectors, the multiplexed configuration, even with
just a few detectors, provides better performance than the other
configurations. In fact, only when the deadtime of single detector
is made shorter than the pulse separation time, can it achieve the
same performance as the multiplexed scheme.

\begin{figure}[tbp]
%[htbp]
\par
\begin{center}
\includegraphics[angle=0, width=15 cm]{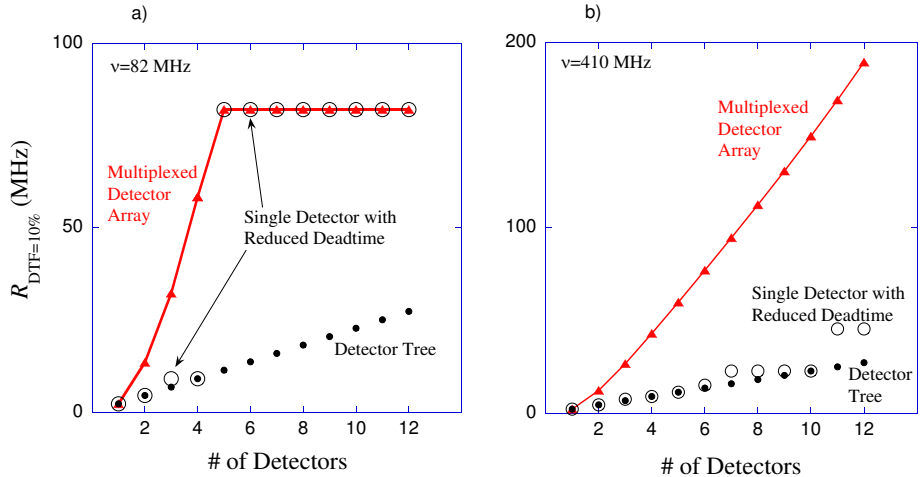}
\end{center}
\caption{$R_{DTF=10\%}$ (the incident rate at which the DTF reaches
10$\%$) of a multiplexed detector array (triangles), a single
detector with reduced deadtime (open circles), and a detector tree
configuration (solid circles) for pulsed sources with repetition
rates of (a) 82 MHz, (b) 410 MHz. The single detector points in (a)
and (b) are plotted versus the 1/\emph{N} reduction in
$T_\text{d}$.} \label{Figure 6}
\end{figure}

\section{Conclusion}
We have shown that a pool of \emph{N} detectors with a controlled
switch system can in principle be operated at much higher incident
photon rates than is otherwise possible either with a single
detector with much reduced deadtime, or an array of detectors with
passive switch system such as might be implemented with by a tree of
beamsplitters \footnote{This latter result is not too surprising as
the schemes using beamsplitter trees and detector arrays were
created to overcome the lack of PNR capability of most
photon-counting detectors \cite{KOB01, RHH03, PTK96}. In other
words, they are designed to solve the problem of photons arriving at
exactly the same time rather than just arriving very close in time.
While our scheme enhances $R_{\text{DTF}=10\%}$, it provides no PNR
capability.}. This advantage holds for both CW and pulsed sources.
 We note from a practical view, that a multiplexed system may be easier
 to implement for a pulsed source as the switching time need only be smaller than the
time separation between pulses. We also note that two factors are
working to increase the relevance of this scheme - a) advancing
quantum information applications are increasing the need for higher
performance detectors, and b) improving array detectors and low-loss
high-speed switches are making this scheme more practical. Moreover
our scheme could also be implemented with photon-number-resolving
(PNR) detectors, as well as the non-photon-number-resolving
detectors typically used for ``photon counting" and analyzed in this
work. The advantage of reduced deadtime, in combination with a PNR
detector array, would make for a very powerful detection capability
indeed.

This work was supported in part by DTO, ARO, and DARPA/QUIST.

\appendix
\section{Pulsed process and geometric distribution}
In analogy with Ref. \cite{renewal}, where the connection between
the Poissonian process and the Poissonian probability distribution
are described, we describe the connection between the pulsed process
and the geometric distribution. Consider a pulsed process, with
probability $p$ of detecting an event for each pulse. The
probability of waiting $n$ pulses before detecting an event is given
by the geometric probability distribution
\begin{equation} \nonumber
\mathcal{T}(n)=p(1-p)^{n-1}.
\end{equation}

The probability of waiting $n$ pulses before detecting two events
(meaning that the second event is detected at the $n^{th}$ pulse) is
\begin{equation} \nonumber
\mathcal{T}_{2}(n)=p B(1|n-1,p),
\end{equation}
and analogous arguments hold for the probability of waiting $n$
pulses before detecting three, four, etc... events. Thus in general
the probability of waiting $n$ pulses before detecting $k-1$ events
is given by the generalized geometric probability
\begin{equation} \nonumber
\mathcal{T}_{k}(n)=p B(k-1|n-1,p).
\end{equation}
Thus the probability that there are more than $k$ events in
$\mathcal{N}$ pulses is $P(m\geq k ~ \mathrm{int} ~ \mathcal{N})=
\sum_{n=k}^{\mathcal{N}}\mathcal{T}_{k}(n)$. The probability of
exactly $k$ events in $\mathcal{N}$ pulses obviously is
\begin{equation} \nonumber
P(m\geq k ~ \mathrm{in} ~ \mathcal{N})-P(m\geq k+1 ~ \mathrm{in} ~
\mathcal{N})=\sum_{n=k}^{\mathcal{N}}\mathcal{T}_{k}(n)-\sum_{n=k+1}^{\mathcal{N}}\mathcal{T}_{k+1}(n).
\end{equation}
Ultimately we see that
$\sum_{n=k}^{\mathcal{N}}\mathcal{T}_{k}(n)-\sum_{n=k+1}^{\mathcal{N}}\mathcal{T}_{k+1}(n)=B(k|\mathcal{N},p)$.

\end{document}